\documentstyle[12pt]{article}                    
\newfont{\ffont}{msym10}                          
\newcommand{\beq}{\begin{equation}}               
\newcommand{\eeq}{\end{equation}}                 
\newcommand{\bqry}{\begin{eqnarray}}              
\newcommand{\eqry}{\end{eqnarray}}                
\newcommand{\bqryn}{\begin{eqnarray*}}            
\newcommand{\eqryn}{\end{eqnarray*}}              
\newcommand{\NL}{\nonumber \\}                    
\newcommand{\PD}[2]                               
    {\frac{\partial^{#2}}{\partial #1^{#2}}}      
\begin{document}
\title{Gell-Mann--Okubo Mass Formula Revisited}
\author{\\ L. Burakovsky\thanks{E-mail: BURAKOV@PION.LANL.GOV} \
and \ T. Goldman\thanks{E-mail: GOLDMAN@T5.LANL.GOV} \
\\  \\  Theoretical Division, MS B285 \\  Los Alamos National Laboratory \\ 
Los Alamos, NM 87545, USA \\}
\date{ }
\maketitle
\begin{abstract}
We show that the application of Regge phenomenology to SU(4) meson multiplets
leads to a new Gell-Mann--Okubo mass-mixing angle formula in the SU(3) sector,
$3m_1^2+\cos ^2\theta \;\!m_0^2+\sin ^2\theta \;\!m_{0^{'}}^2+\sqrt{2}\sin 2
\theta \left( m_0^2-m_{0^{'}}^2\right) =4m_{1/2}^2,$ where $m_1,m_{1/2},m_0,m
_{0^{'}}$ are the masses of the isovector, isodoublet, isoscalar mostly octet 
and isoscalar mostly singlet states, respectively, and $\theta $ is the nonet 
mixing angle. For an ideally mixed nonet, $\theta =\arctan \;\!1/\sqrt{2},$ 
this formula reduces to $2m^2_0+3m^2_1=4m^2_{1/2}+m^2_{0^{'}}$ which holds 
with an accuracy of $\sim 1$\% for vector and tensor mesons. For pseudoscalar 
mesons, with the $\eta $-$\eta ^{'}$ mixing angle $-\arctan \;\!1/(2\sqrt{2})
\simeq -19.5^o,$ in agreement with experiment, it leads to the relation $4m_K^
2=3m_\pi ^2+m_{\eta ^{'}}^2$ which holds to an accuracy of better than 1\% 
for the measured pseudoscalar meson masses.
\end{abstract}

\section*{ }
The original Gell-Mann--Okubo (GMO) mass formula \cite{GMO}
\beq
m_1^2+3m_8^2=4m_{1/2}^2
\eeq
relates the masses of the isovector $(m_1),$ isodoublet $(m_{1/2})$ and
isoscalar octet $(m_8)$ states of a meson octet. It is usually cast into a
form which relates $m_1$ and $m_{1/2}$ with the masses of the two physical 
isoscalar states, $m_{0^{'}}$ and $m_0,$ which have $n\bar{n}$ and $s\bar{s}$
quark content, respectively $(n$ stands for non-strange $u$- and $d$-quark),
assuming the ideal structure of a nonet:
\beq
\frac{m_1^2+m_{0^{'}}^2}{2}+m_0^2=2m_{1/2}^2.
\eeq 
Indeed, in general the isoscalar octet $(\omega _8)$ and singlet $(\omega _9)$
states get mixed, because of SU(3) breaking, which results in the physical
$\omega _0$ and $\omega _{0^{'}}$ states (the $\omega _0$ is a mostly octet 
isoscalar): $$\omega _0=\omega _8\cos \theta -\omega _9\sin \theta ,$$ 
$$\omega _{0^{'}}=\omega _8\sin \theta +\omega _9\cos \theta ,$$ where 
$\theta $ is the mixing angle. Assuming, as usual, that the relevant matrix 
elements are equal to the squared masses of the corresponding states, one 
obtains from the above relations \cite{Per}
\beq
m_0^2=m_8^2\cos ^2\theta +m_9^2\sin ^2\theta -2m_{89}^2\sin \theta \cos 
\theta ,
\eeq
\beq
m_{0^{'}}^2=m_8^2\sin ^2\theta +m_9^2\cos ^2\theta +2m_{89}^2\sin \theta \cos 
\theta .
\eeq
Since $\omega _0$ and $\omega _{0^{'}}$ as physical states are orthogonal, one
has further
\beq
m_{00^{'}}^2=0=(m_8^2-m_9^2)\sin \theta \cos \theta +m_{89}^2(\cos ^2\theta -
\sin ^2\theta ).
\eeq
Eliminating $m_9$ and $m_{89}$ from (3)-(5) yields
\beq
\tan ^2\theta =\frac{m_8^2-m_0^2}{m_{0^{'}}^2-m_8^2}.
\eeq
It also follows from (3)-(5) that $m_8^2=m_0^2 \cos ^2 \theta +m_{0^{'}}^2
\sin ^2\theta ,$ and therefore, Eq. (1) may be rewritten as
\beq
4m_{1/2}^2-m_1^2-3m_0^2=3\left( m_{0^{'}}^2-m_0^2\right) \sin ^2\theta ,
\eeq
which is the Sakurai mass formula \cite{Sakurai}. For the ideal 
octet-singlet mixing, $\tan \theta _{id}=1/\sqrt{2},$ for which $\omega _0=
s\bar{s},$ $\omega _{0^{'}}=(u\bar{u}+d\bar{d})/\sqrt{2},$ Eq. (7) reduces to 
(2), the formula for the ideal structure of a nonet.

The formula (2) is known to hold with a high accuracy for all well established 
meson multiplets except for the pseudoscalar one. It is widely believed that 
the reason for the invalidity of Eq. (2) for pseudoscalar mesons is a large
dynamical mass of the isoscalar singlet state developed (before its mixing with
the isoscalar octet which results in the physical $\eta $ and $\eta ^{'}$
states) due to axial $U(1)$ symmetry breakdown. In fact, the observed mass
splitting among the pseudoscalar nonet may be induced by the following symmetry
breaking terms \cite{symbr},
\beq
L^{(0)}_m=\frac{f^2}{4}\left( r\;Tr\;\!M\left( U+U^{+}\right) +\frac{\alpha }{
4N}\left[ \;Tr\left( \ln U-\ln U^{+}\right) \right] ^2\right) ,\;\;\;r={\rm 
const,}
\eeq
with $M$ being the quark mass matrix, 
\beq
M={\rm diag}\;(m_u,\;m_d,\;m_s)=m_s\;{\rm diag}\;(x,y,1),\;\;\;x\equiv \frac{
m_u}{m_s},\;y\equiv \frac{m_d}{m_s},
\eeq
in addition to the U(3)$_L\times $U(3)$_R$ invariant non-linear Lagrangian 
\beq
L^{(0)}=\frac{f^2}{4}\;Tr\;\!\left( \partial _\mu U\partial ^\mu U\right) ,
\eeq
with $$U=\exp (i\pi /f) ,\;\;\;\pi \equiv \lambda _a\pi ^a,\;\;a=0,1,\ldots ,
8,$$ which incorporates the constraints of current algebra for the light 
pseudoscalars $\pi ^a$ \cite{Georgi}. As pointed out in ref. \cite{KM}, chiral
corrections can be important, the kaon mass being half the typical 1 GeV chiral
symmetry breaking scale. Such large corrections are precisely required from 
the study of the octet-singlet mass squared matrix $M^2.$ In the isospin limit
$x=y$ one has
\beq
M^2=\frac{1}{3}\left(
\begin{array}{cc}
4m_K^2\;-\;m_\pi ^2 & -2\sqrt{2}(m_K^2-m_\pi ^2) \\
-2\sqrt{2}(m_K^2-m_\pi ^2) & 2m_K^2+m_\pi ^2+3\alpha 
\end{array}
\right) .
\eeq
The $\eta $-$\eta ^{'}$ mixing angle $\theta _{\eta \eta ^{'}}$ then reads
\beq
\tan 2\theta _{\eta \eta ^{'}}=\frac{2m_{89}^2}{m_{9}^2-m_{8}^2}=
2\sqrt{2}\left( 1-\frac{3\alpha }{2(m_K^2-m_\pi ^2)}\right) ^{-1}.
\eeq
For $\alpha =0,$ one obtains the ideal mixing and the mass relation $m_{\eta ^{
'}}=m_\pi $ as the source of the U(1) problem \cite{Wei}. The parameter 
$\alpha $ is assumed to be induced by instantons \cite{symbr,Dmitra} and is 
determined by the trace condition $\alpha =m_\eta ^2+m_{\eta ^{'}}^2-2m_K^2
\simeq 0.725$ GeV$^2;$ one then obtains form (12) the mixing angle
\beq
\theta _{\eta \eta ^{'}}\simeq -18.5^o,
\eeq
in agreement with most of experimental data \cite{data}. A more popular way
to extract the $\eta $-$\eta ^{'}$ mixing angle, through the relation (6)
based on GMO (1) \cite{pdg}, leads to $$\theta _{\eta \eta ^{'}}\simeq -10.5^
o,$$ in disagreement with experiment \cite{data}. On the other hand, in the 
octet approximation $m_8\approx m_\eta ,$ GMO (1) is thought to be quite 
successful since it predicts $$m_8\approx 567\;{\rm MeV,}$$ which is within 
the physical $\eta $ mass of 547.5 MeV with an accuracy of $\sim $ 3.5\%. 
Thus, GMO for pseudoscalar mesons is believed to be
\beq
4m_K^2=3m_\eta ^2+m_\pi ^2,
\eeq
and although it does not reproduce the $\eta $-$\eta ^{'}$ mixing angle 
correctly, it gives the mass of the $\eta $ with a rather high accuracy.

In fact, the octet approximation and the corresponding relation (14) mean
$$\eta \approx \eta _8=\frac{u\bar{u}+d\bar{d}-2s\bar{s}}{\sqrt{6}},$$
consistent with the $(1/3\;n,\;2/3\;s)$ quark content of the $\eta ,$ in 
agreement with the Gell-Mann--Oakes-Renner relations (to first order in chiral
symmetry breaking) \cite{GOR}
\bqry
m_\pi ^2 & = & 2\;\!B\;m_n, \NL
m_K^2 & = & B\;(m_s+m_n), \NL
m_\eta ^2 & = & \frac{2}{3}\;\!B\;(2m_s+m_n),
\eqry
$$m_n\equiv \frac{m_u+m_d}{2};\;\;\;B={\rm const;}$$
however, the actual quark content of the $\eta ,$ due to the $\eta _8$-$\eta _
9$ mixing angle $\simeq -19^o,$ is \cite{DGH}   
\beq
\eta \;\simeq \;0.58\;(u\bar{u}+d\bar{d})-0.57\;s\bar{s}\;\approx \;
\frac{u\bar{u}+d\bar{d}-s\bar{s}}{\sqrt{3}},
\eeq
i.e., $(2/3\;n,\;1/3\;s),$ quite different from that provided by (14). Thus,
a natural suspicion is that $m_8\approx m_\eta $ is purely numerical 
coincidence, and the actual Gell-Mann--Okubo relation should be different from
Eq. (14). Moreover, since the $(1/3\;n,\;2/3\;s)$ quark content corresponds to
the $\eta ^{'}$ meson, in view of \cite{DGH}
\beq
\eta ^{'}\;\simeq \;0.40\;(u\bar{u}+d\bar{d})+0.82\;s\bar{s}\;\approx \; 
\frac{u\bar{u}+d\bar{d}+2s\bar{s}}{\sqrt{6}},
\eeq
one may also suspect that the true mass formula should relate the masses of the
$\pi ,$ $K$ and $\eta ^{'}.$ 

In this paper we derive such a formula which we call the Gell-Mann--Okubo mass
formula revisited (GMO$_r).$ We shall use Regge phenomenology which proved to 
be quite successful in producing hadronic (both mesonic \cite{mes,quadr} and 
baryonic \cite{bar}) mass relations in both the light and heavy quark sectors. 

As discussed in detail in our previous papers \cite{mes,quadr,bar}, Regge
phenomenology for mesons is based on the following two relations among the 
masses and Regge slopes of the states which belong to a given meson mulitplet:
\bqry
\alpha ^{'}_{i\bar{i}}m^2_{i\bar{i}}\;+\;\alpha ^{'}_{j\bar{j}}m^2_{j\bar{j}} 
 & = & 2\alpha ^{'}_{j\bar{i}}m^2_{j\bar{i}}, \\  
\frac{1}{\alpha ^{'}_{i\bar{i}}}\;+\;\frac{1}{\alpha ^{'}_{j\bar{j}}} & = & 
\frac{2}{\alpha ^{'}_{j\bar{i}}},
\eqry
In the light quark sector, $i=n\;\!(\;\!=u,d\;\!),\;j=s,$ one has $\alpha ^{'}
_{s\bar{s}}\approx \alpha ^{'}_{s\bar{n}}\approx \alpha ^{'}_{n\bar{n}};$ with
the definition
\beq
m_{n\bar{n}}^2\equiv \frac{m^2_{n\bar{n}}(I=1)+m^2_{n\bar{n}}(I=0)}{2}
\eeq
$(I$ stands for isospin), Eq. (18) then reduces to the formula (2).

We shall also use the following mass relations among vector and pseudoscalar 
meson mass squared: 
\bqry
 &   & \;m_\rho ^2\;-\;m_\pi ^2\;\;\approx \;\;m_{K^\ast }^2\;-\;m_K^2 \NL
 & \approx  & m_{D^\ast }^2\;-\;m_D^2\;\approx \;m_{D_s^\ast }^2\;-\;m_{D_s}^2
\;\simeq \;0.57\;{\rm GeV}^2.
\eqry
This relations are easily obtained in the constituent quark model \cite{LSG} 
and an algebraic approach to QCD \cite{OT}. They lead, e.g., to
\beq
m_\phi ^2-\frac{m_\rho ^2+m_\omega ^2}{2}=2\left(m_{K^\ast }^2-\frac{m_\rho ^2
+m_\omega ^2}{2}\right) \approx 2\left( m_K^2-m_\pi ^2\right) ,
\eeq
in view of (2) and $m_\omega \approx m_\rho .$

It is easily seen that the relations (18),(19) may be applied only to pure 
$q\bar{q}$ states, and neither $\eta $ nor $\eta ^{'}$ is such a state. 
Therefore, in order to apply Eqs. (18),(19) to pseudoscalar mesons, one has
first to construct the proper states $\eta _n$ and $\eta _s$ (as linear 
combinations of the physical $\eta $ and $\eta ^{'}),$  
\beq
\eta _n=\frac{u\bar{u}+d\bar{d}}{\sqrt{2}},\;\;\;\eta _s=s\bar{s},
\eeq
which have the masses $m_{\eta _n}$ and $m_{\eta _s},$ respectively, which we
determine later on. For these states, we apply Eqs. (18),(19) with $(i,j)=(n,
c)$ and $(i,j=s,c)$ using the experimental fact that the slope of the $c\bar{c
}$-trajectory is less than that of the $n\bar{n}$- and $s\bar{s}$-trajectories.
One has, in agreement with (19),
\beq
\alpha ^{'}_{c\bar{s}}\approx \alpha ^{'}_{c\bar{n}}=\frac{\alpha ^{'}_{n\bar{
n}}}{1+x}\approx \frac{\alpha ^{'}_{s\bar{s}}}{1+x},\;\;\alpha ^{'}_{c\bar{c}}
=\frac{\alpha ^{'}_{n\bar{n}}}{1+2x}\approx \frac{\alpha ^{'}_{s\bar{s}}}{1+
2x},\;\;x>0.
\eeq
Therefore,
\bqry
m^2_{n\bar{n}}\;+\;\frac{m^2_{c\bar{c}}}{1+2x} & = & 2\frac{m^2_{c\bar{n}}}{1+
x}, \\
m^2_{s\bar{s}}\;+\;\frac{m^2_{c\bar{c}}}{1+2x} & = & 2\frac{m^2_{c\bar{s}}}{1+
x},
\eqry
with $m^2_{n\bar{n}}$ defined in (20), in agreement with Eq. (2) in the light
quark sector.

For pseudoscalar and vector mesons, Eqs. (25),(26) may be rewritten as
\bqry
m_{\tilde{\eta }_n}^2\;+\;\frac{m_{\eta _c}^2}{1+2x} & = & 2\frac{m_D^2}{1+x},
 \NL
m_{\eta _s}^2\;+\;\frac{m_{\eta _c}^2}{1+2x} & = & 2\frac{m_{D_s}^2}{1+x}, \NL
m_{\tilde{\rho }}^2\;+\;\frac{m_{J/\psi }^2}{1+2x} & = & 2\frac{m_{D^\ast }^
2}{1+x}, \NL
m_{\phi }^2\;+\;\frac{m_{J/\psi }^2}{1+2x} & = & 2\frac{m_{D^\ast _s}^2}{1+x},
\eqry 
where 
\beq
m^2_{\tilde{\eta }_n}\equiv \frac{m^2_\pi +m^2_{\eta _n}}{2},\;\;\;
m^2_{\tilde{\rho }}\equiv \frac{m^2_\rho +m^2_\omega }{2}.
\eeq
It follows from (21),(22),(27) that
\bqry
m^2_{\eta _s}\;-\;m^2_{\tilde{\eta }_n} & = & \frac{2}{1+x}\left( m^2_{D_s}\;-
\;m_D^2\right) \;\approx \;\frac{2}{1+x}\left( m^2_{D^\ast _s}\;-\;m^2_{D^\ast
}\right)   \NL
 & = & m^2_\phi \;-\;m^2_{\tilde{\rho }}\;\approx \;2\left( m_K^2\;-\;m_\pi ^2
\right) .
\eqry 
One then obtains, from (28),(29),
\beq
2m^2_{\eta _s}-m^2_{\eta _n}\cong 4m_K^2-3m_\pi ^2,
\eeq
which is a new Gell-Mann--Okubo (GMO$_r)$ mass formula for pseudoscalar mesons,
which however cannot be applied to them directly since $m_{\eta _n}$ and $m_{
\eta _s}$ are not known. Therefore, the last step in derivation of the analog 
of Eq. (30) applicable to pseudoscalar states is to determine the values of $m_
{\eta _n}$ and $m_{\eta _s},$ in terms of the physical $m_\eta $ and $m_{\eta 
^{'}},$ in order to use them in Eq. (30).  

We assume the $\eta $-$\eta ^{'}$ mixing angle to take the ``ideal'' value
\beq
\theta _{\eta \eta ^{'}}=-\arctan \frac{1}{2\sqrt{2}}\approx -19.5^o,
\eeq
in agreement with experimental data \cite{data}. This value was first predicted
by Bramon \cite{Bra} from a simple quark model and duality constraints for the 
set of scattering processes $\pi \eta \rightarrow \pi \eta ,$ $\pi \eta
\rightarrow \pi \eta ^{'},$ $\pi \eta ^{'}\rightarrow \pi \eta ^{'},$
$\eta K\rightarrow (\pi ,\eta ,\eta ^{'})K,$ and $\eta \eta \rightarrow \eta 
\eta ,$ $\eta \eta \rightarrow \eta \eta ^{'},$ $\eta \eta ^{'}\rightarrow
\eta \eta ^{'}.$\footnote{The mixing angle (31) predicts the suppression of the
$K^{\ast }_2\rightarrow K\eta $ decay \cite{Bra}, in excellent agreement with 
experiment \cite{pdg}.} Since the ideal mixing of a nonet corresponds to 
$\theta _{id}=\arctan \;\!1/\sqrt{2}\approx 35.3^o,$ one has from (31)
\beq
2\theta _{id}-\theta _{\eta \eta ^{'}}=\frac{\pi }{2}.
\eeq
In view of (32) and
$$\left(
\begin{array}{c}
\eta  \\
\eta ^{'}
\end{array}
\right) =\left(
\begin{array}{lr}
\cos \theta _{\eta \eta ^{'}} & -\sin \theta _{\eta \eta ^{'}} \\
\sin \theta _{\eta \eta ^{'}} & \cos \theta _{\eta \eta ^{'}}
\end{array}
\right) \left(
\begin{array}{c}
\eta _8 \\
\eta _9
\end{array}
\right) ,\;\;\left(
\begin{array}{c}
\eta _s \\
\eta _n
\end{array}
\right) =\left(
\begin{array}{lr}
\cos \theta _{id} & -\sin \theta _{id} \\
\sin \theta _{id} & \cos \theta _{id}
\end{array}
\right) \left(
\begin{array}{c}
\eta _8 \\
\eta _9
\end{array}
\right) ,$$ one has
$$\left(
\begin{array}{c}
\eta _s \\
\eta _n
\end{array}
\right) =\left(
\begin{array}{lr}
\cos (\theta _{id}-\theta _{\eta \eta ^{'}}) & -\sin (\theta _{id}-\theta _{
\eta \eta ^{'}}) \\
\sin (\theta _{id}-\theta _{\eta \eta ^{'}}) & \cos (\theta _{id}-\theta _{
\eta \eta ^{'}})
\end{array}
\right) \left(
\begin{array}{c}
\eta  \\
\eta ^{'}
\end{array}
\right) $$
\beq
=\left(
\begin{array}{lr}
\sin \theta _{id} & -\cos \theta _{id} \\
\cos \theta _{id} & \sin \theta _{id}
\end{array}
\right) \left(
\begin{array}{c}
\eta  \\
\eta ^{'}
\end{array}
\right) .
\eeq
Assuming, as previously, that the relevant matrix elements are equal to the 
squared masses of the corresponding states, and using the orthogonality of the
$\eta $ and $\eta ^{'}$ as physical states, we obtain
\bqry
m^2_{\eta _n} & = & \frac{2}{3}\;m^2_{\eta }\;+\;\frac{1}{3}\;m^2_{\eta ^{'}},
 \\
m^2_{\eta _s} & = & \frac{1}{3}\;m^2_{\eta }\;+\;\frac{2}{3}\;m^2_{\eta ^{'}},
\eqry
in agreement with naive expectations from the quark content of these states.
Thus, $2m^2_{\eta _s}-m^2_{\eta _n}=m^2_{\eta ^{'}};$ the use of this result 
in Eq. (30) leads finally to 
\beq
4m_K^2\cong 3m_\pi ^2+m_{\eta ^{'}}^2,
\eeq
which is our final form of GMO$_r$ for pseudoscalar mesons.

With the measured masses of the states entering Eq. (36) \cite{pdg},
$$m_\pi =137.3\pm 2.3\;{\rm MeV,}\;\;\;m_K=495.7\pm 2.0\;{\rm MeV,}\;\;\;
m_{\eta ^{'}}=957.8\;{\rm MeV,}$$ it gives (in GeV$^2)$ $0.983\pm 0.008$ on 
the l.h.s. vs. $0.974\pm 0.002$ on the r.h.s.; its accuracy is therefore 
$\simeq $ 0.9\%.
 
The value of $m^2_{\eta _8}$ may be obtained from (6),(31):
$$\frac{m^2_{\eta _8}-m^2_\eta }{m^2_{\eta ^{'}}-m^2_{\eta _8}}=\frac{1}{8};$$
therefore
\beq
9m^2_{\eta _8}=8m^2_{\eta }+m^2_{\eta ^{'}},\;\;{\rm and}\;\;m_{\eta _8}\approx
607\;{\rm MeV,}
\eeq
in excellent agreement with 
\beq
m^2_{\eta _8}=\frac{4}{3}m_K^2-\frac{1}{3}m_\pi ^2-\frac{2}{3}\frac{m_K^4}{(4
\pi f_\pi )^2}\ln \frac{m_K^2}{\mu ^2}\approx 610\;{\rm MeV,}\;\;\;\mu \approx
1\;{\rm GeV}
\eeq
obtained from chiral perturbation theory \cite{chi}.

We now wish to extend the application of this new Gell-Mann--Okubo mass 
formula (30), which we rewrite here as $$2m^2_{s\bar{s}}+3m^2_{n\bar{n}}(I=1)=
4m^2_{s\bar{n}}+m^2_{n\bar{n}}(I=0).$$ In the case of the ideal mixing of a 
nonet, it differs from (2) only by a term proportional to explicit isospin
variation: $m^2_{n\bar{n}}(I=1)-m^2_{n\bar{n}}(I=0).$ However, in contrast to
(2), it has correct (3 and 1, respectively) isospin degeneracies for the 
isovector and non-strange isoscalar states. Moreover, as we have shown in the 
example of pseudoscalar mesons, this formula may work when its counterpart (2)
does not; namely, in case of a non-ideal nonet mixing. As clear from its 
derivation, the formula (2) will hold only for an ideally mixed nonet; in 
contrast, Eq. (30) is obtained from Regge phenomenology based on Eqs. 
(18),(19) which relate the masses of pure $q\bar{q}$ states but not the nonet 
mixing angle, and will therefore hold even if a nonet mixing differs from the 
ideal one (e.g., if the isoscalar octet mass is shifted from its ``GMO value''
(1), Eq. (6) will not be compatible with $\theta =\theta _{id},$ even if 
$\omega _0=s\bar{s},$ $\omega _{0^{'}}=n\bar{n}).$ The generality of Eqs. 
(18),(19) which are the basis of the formula (30), and of the arguments used 
for its derivation suggests that this formula should be applicable to any 
meson multiplet, not only to the pseudoscalar one. Also, as discussed above, 
in cases when the nonet mixing is not ideal but two isoscalars are almost pure
$n\bar{n}$ and $s\bar{s}$ states (which are realized in the real world in some
cases), we expect this formula to hold with better accuracy than Eq. (2).  

In order to test this, we shall apply the formula (30) to two well-established
meson multiplets, vector and tensor mesons.

For vector mesons, we write down this formula (assuming $\omega \approx n\bar{
n},$ $\phi \approx s\bar{s})$ as
\beq
2m_\phi ^2+3m_\rho ^2=4m_{K^\ast }^2+m_\omega ^2.
\eeq
For the measured meson masses entering Eq. (39), it gives (in GeV$^2)$ 3.85
on the l.h.s. vs $3.81\pm 0.02$ on the r.h.s.; the accuracy is therefore 1.1\%.
For comparison, Eq. (2) for vector mesons gives (in GeV$^2)$ 1.64 vs. $1.60\pm 
0.02,$ with the accuracy of 2.5\%, a factor of two worse than that of (39).

For tensor mesons, Eq. (30) should be written down (with $f_2\approx n\bar{
n},$ $f_2^{'}\approx s\bar{s})$ as
\beq
2m^2_{f_2^{'}}+3m_{a_2}^2=4m_{K_2^\ast }^2+m^2_{f_2},
\eeq
which for the measured meson masses gives (in GeV$^2)$ $9.86\pm 0.03$ on the 
l.h.s. vs. $9.79\pm 0.07$ on the r.h.s., with the accuracy of 0.7\%. Eq. (2)
in this case gives (in GeV$^2)$ $4.01\pm 0.02$ vs. $4.08\pm 0.03,$ with the 
accuracy of 1.7\%, again, a factor of two worse than that of (40).

Finally, we cast Eq. (30) into a form which involves the physical meson masses
and nonet mixing angle, and therefore is applicable to every meson nonet.

It follows from (33) that 
\bqry
m_n^2 & = & \sin ^2\xi \;m_0^2\;+\;\cos ^2\xi \;m_{0^{'}}^2, \NL
m_s^2 & = & \cos ^2\xi \;m_0^2\;+\;\sin ^2\xi \;m_{0^{'}}^2,
\eqry
where $m_0,\;m_{0^{'}}$ are the masses of the isoscalar mostly octet and 
singlet states, respectively, and $\xi \equiv \theta _{id}-\theta ,$ $\theta $
being the nonet mixing angle. Using these expressions for $m^2_n$ and $m^2_s$ 
in Eq. (30), one obtains
\beq
\left( 3\cos ^2\xi -1\right) m_0^2+\left( 2-3\cos ^2\xi \right) m_{0^{'}}^2
=4m_{1/2}^2-3m_1^2,
\eeq
where $m_1,\;m_{1/2}$ are the masses of the isovector and isodoublet states, 
respectively, which finally reduces, through
\beq
\cos \xi =\frac{\sqrt{2}\cos \theta +\sin \theta }{\sqrt{3}},
\eeq
to
\beq
3m_1^2+\cos ^2\theta \;m_0^2+\sin ^2\theta \;m_{0^{'}}^2+\sqrt{2}\sin 2\theta 
\left( m_0^2-m_{0^{'}}^2\right) =4m_{1/2}^2,
\eeq
which is a new nonet mass-mixing angle relation. For an ideally mixed nonet,
$\tan \theta =1/\sqrt{2},$ it reduces to 
$$2m_0^2+3m_1^2=m_{0^{'}}^2+4m_{1/2}^2,$$ which is equivalent to (30); for the
pseudoscalar nonet, $\tan \theta =-1/(2\sqrt{2}),$ it leads to
$$3m_1^2=4m_{1/2}^2-m_{0^{'}}^2,$$ which is equivalent to (36).

The new mass-mixing angle relation (44) is the main result of this paper.  

\section*{Concluding remarks}
We have derived a new Gell-Mann--Okubo mass formula, Eq. (30), by applying 
Regge phenomenology to pseudoscalar and vector mesons. For pseudoscalar mesons,
using the $\eta $-$\eta ^{'}$ mixing angle $\simeq -19.5^o,$ in agreement with
experiment, this formula may be reduced to Eq. (36) which relates the masses
of the $\pi ,$ $K$ and $\eta ^{'}$ mesons. This relation predicts the mass of 
the $\eta ^{'}$ meson, $m_{\eta ^{'}}=962.4\pm 5.1$ MeV, within the physical 
mass of 957.8 MeV with an accuracy of $\simeq $ 0.5\%. Since no additional 
assumption except the linearity of the corresponding trajectories and the 
additivity of the inverse slopes, Eq. (19), has been made in deriving Eqs. (30)
and (36) (Eq. (36) is based on the $\eta $-$\eta ^{'}$ mixing angle $\simeq 
-19.5^o$ which is provided by duality constraints \cite{Bra}), we conclude that
Regge phenomenology suffices to describe the $\eta ^{'}$ mass, which has been
a mystery for a long time. The question remains, however, about the mass of the
isoscalar singlet state (before its mixing with the isoscalar octet which 
results in the physical $\eta $ and $\eta ^{'}$ states): since, independent of
the mixing angle, $m_{\eta _8}^2+m_{\eta _9}^2=m_\eta ^2+m_{\eta ^{'}}^2$ (as
seen, e.g., in Eqs. (3),(4)), the value of $m_{\eta _8}$ (37) leads to $m_{
\eta _9}\approx 921$ MeV. Thus, compared to a 40 MeV shift of the mass of the 
$\eta _8$ from its GMO value by chiral corrections, which is only 7\% of its
bare (GMO) mass, the mass of the $\eta _9$ is shifted by $\sim 500$ MeV, 
taking its ``GMO'' value as $\simeq (2m_K^2-m_{\eta _8}^2)^{1/2}.$ We believe 
that such a large shift of the mass of the pseudoscalar isoscalar octet state 
is due to instanton effects discussed in detail in refs. \cite{Dmitra,inst}.  

It is clear from our arguments given above that Eq. (30) or its mass-mixing 
angle form (44) should also hold for scalar mesons. It would be very 
interesting to consider the scalar meson case and shed some light on the 
long-standing problem of the correct $q\bar{q}$ assignment for this nonet. We 
plan to do this in a forthcoming publication.
 
Also, the generalization of the relations for meson masses and mixing angles 
discussed in the paper to finite temperature and/or baryon density would be 
very important for the understanding of the in-medium hadron behavior and its 
possible consequences for the decay widths and particle spectra, in view of 
ongoing experimental activity of different groups all around the world in the 
search for the new phases of matter.  

\section*{Acknowledgements} 
Correspondence of one of the authors (L.B.) with L.P. Horwitz during the 
preparation of the present paper is greatly acknowledged.

\end{document}